\documentclass[conf]{IEEEtran}
\usepackage{algorithm,algpseudocode}
\usepackage{amsmath}
\usepackage{amssymb}
\usepackage{mathtools}

\usepackage{cite}
\def\BibTeX{{\rm B\kern-.05em{\sc i\kern-.025em b}\kern-.08em
    T\kern-.1667em\lower.7ex\hbox{E}\kern-.125emX}}
\usepackage[bookmarksnumbered=true]{hyperref} 

\hypersetup{
     colorlinks = true,
     linkcolor = blue,
     anchorcolor = blue,
     citecolor = blue,
     filecolor = blue,
     urlcolor = blue
     }

\usepackage{svg}

\usepackage{wrapfig}
\usepackage{caption}
\usepackage{subcaption}
\usepackage{booktabs}

\usepackage{xcolor}
\begin{document}

\title{Low-power Spike-based Wearable Analytics on RRAM Crossbars}

\author{
\IEEEauthorblockN{Abhiroop Bhattacharjee$^{*\$}$, Jinquan Shi$^{*\$}$, Wei-Chen Chen$^\circ$, Xinxin Wang$^\circ$ and Priyadarshini Panda$^\$$} \\

$^\$$Department of Electrical Engineering, Yale University, USA \\
$^\circ$ Department of Electrical Engineering, Stanford University, USA

\thanks{$^*$ These authors have contributed equally to this work.\\ 
This work was supported in part by CoCoSys, a JUMP2.0 center sponsored by DARPA and SRC, the National Science Foundation (CAREER Award, Grant \#2312366, Grant \#2318152), the DoE MMICC center SEA-CROGS (Award \#DE-SC0023198), and the National Science Foundation Expeditions in Computing (Penn State, Award \#1317470). The authors acknowledge Weier Wan and Prof. H.-S. Philip Wong for their valuable discussions and insights towards this work.}
}

\IEEEaftertitletext{\vspace{-2.5\baselineskip}}
% make the title area
\maketitle

% As a general rule, do not put math, special symbols or citations
% in the abstract or keywords.

\begin{abstract}

This work introduces a spike-based wearable analytics system utilizing Spiking Neural Networks (SNNs) deployed on an In-memory Computing engine based on RRAM crossbars, which are known for their compactness and energy-efficiency. Given the hardware constraints and noise characteristics of the underlying RRAM crossbars, we propose online adaptation of pre-trained SNNs in real-time using Direct Feedback Alignment (DFA) against traditional backpropagation (BP). Direct Feedback Alignment (DFA) learning, that allows layer-parallel gradient computations, acts as a fast, energy \& area-efficient method for online adaptation of SNNs on RRAM crossbars, unleashing better algorithmic performance against those adapted using BP. Through extensive simulations using our in-house hardware evaluation engine called $DFA\_Sim$, we find that DFA achieves upto $64.1\%$ lower energy consumption, $10.1\%$ lower area overhead, and a $2.1\times$ reduction in latency compared to BP, while delivering upto $7.55\%$ higher inference accuracy on human activity recognition (HAR) tasks.

\end{abstract}

% Note that keywords are not normally used for peerreview papers.
\vspace{-4mm}

\IEEEpeerreviewmaketitle

\section{Introduction}
\label{sec:intro}

%The rise of wearable technologies in edge computing scenarios has drawn attention to tasks such as diagnostics, smart healthcare, and fitness monitoring, which often involve real-time time-series data sensing and processing, as shown in Fig. \ref{fig1}(a) \cite{tanzarella2023neuromorphic, bian2023evaluating, andante}. Wearable devices are typically limited to low power budgets of less than $1 W$, particularly for tasks such as human activity recognition (HAR), physiological monitoring, and predictive health diagnostics \cite{covi2021adaptive, furber2014spinnaker, li2023efficient}. Achieving reliable data processing within these low-power constraints is a significant challenge, emphasizing the need for energy-efficiency in edge-based systems.

\begin{figure}[t]
    \centering
    
    \includegraphics[width=0.9\linewidth]{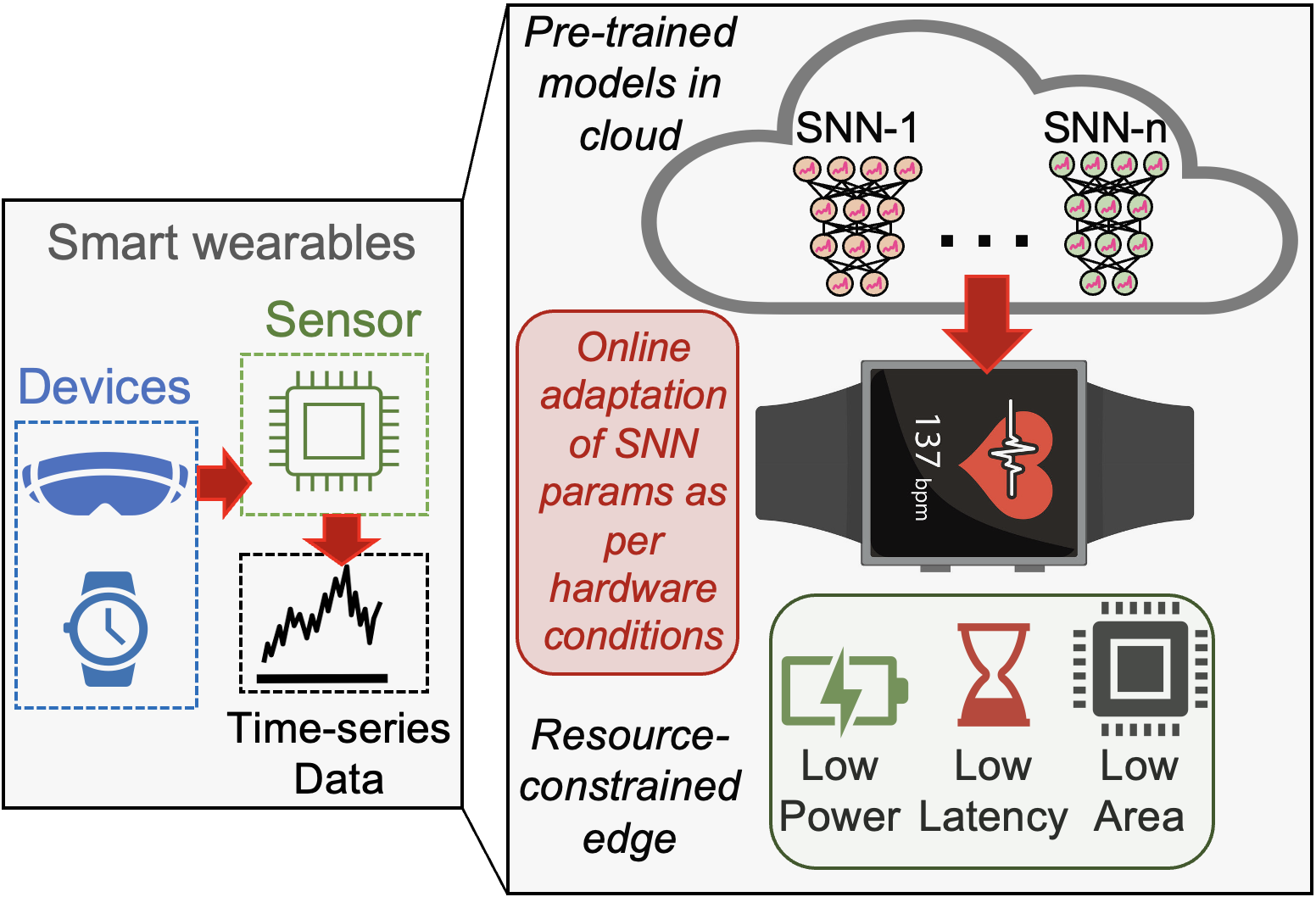}%

    %\captionsetup{justification=centering}
    \caption{Pictorial depiction of SNNs used in wearables for temporal data-processing. Pre-trained SNNs in the cloud are adapted online according to the constraints of resource-constrained edge devices.}
    \label{fig1}

    \vspace{-6mm}
\end{figure}

The rise of wearable technologies in edge computing has drawn attention to tasks such as diagnostics, smart healthcare, and fitness monitoring, which often involve real-time time-series data processing (Fig. \ref{fig1}(a)) \cite{tanzarella2023neuromorphic, bian2023evaluating, andante}. Wearable devices typically operate under low power budgets of less than $1 W$, especially for tasks like human activity recognition (HAR), physiological monitoring, and predictive health diagnostics \cite{covi2021adaptive, furber2014spinnaker, li2023efficient}. This makes energy-efficiency crucial in edge systems. Traditional deep learning models, while powerful, are energy-intensive due to dense matrix multiplications and frequent memory access, making them unsuitable for low-power wearables \cite{jouppi2017datacenter, garcia2019estimation}. Today, Spiking Neural Networks (SNNs) have emerged as an energy-efficient alternative with their sparse \& event-driven binary spike processing, particularly suited for real-time temporal tasks like ECG, EEG, motion tracking, and speech analysis \cite{shaaban2024rt, tanzarella2023neuromorphic, feng2023towards, li2023efficient}.

From a hardware implementation standpoint, In-memory Computing (IMC) with analog crossbar arrays enables compact, energy-efficient dot-product operations with high parallelism \cite{verma2019memory, shanbhag2022comprehending}. Unlike traditional von-Neumann architectures like GPUs and TPUs, IMC crossbars keep neural network's weights stationary, reducing data transfer overhead between the memory and the compute units. This is ideal for wearables, which have stringent area and power constraints. IMC-implemented SNNs, with their high spike sparsity and binary computations, offer reduced peripheral circuit and data communication overhead, enhancing energy-efficiency and throughput \cite{moitra2024when, mehonic2022brain, ankit2017resparc}.

%Despite the potential advantages, IMC-implemented SNNs are vulnerable non-idealities owing to their analog nature of computing dot-products over multiple timesteps \cite{bhattacharjee2022examining, moitra2023spikesim, moitra2024when}. This challenges their widespread deployment to real-world edge applications. The non-idealities stem from the limited precision and read variations of the non-volatile memory (NVM) devices in the crossbars, resulting in inaccurate dot-products and, thereby, reducing inference accuracy of the deployed SNN workloads \cite{sun2019impact, bhattacharjee2022examining}. A prevailing approach to non-ideality mitigation is to perform Variation-aware-Training (VAT) of SNNs on the deployed IMC platform to build robustness against stochastic variations affecting inference \cite{liu2015vortex, chakraborty2018technology}. Therefore, in order to infer an SNN on a resource-constrained edge platform, such as IMC crossbars, online adaptation of the pre-trained model to the conditions of the underlying hardware is imperative (see Fig. \ref{fig1}). 

However, IMC-implemented SNNs are prone to non-idealities due to the analog nature of dot-product operations over multiple timesteps \cite{bhattacharjee2022examining, moitra2023spikesim, moitra2024when}. These non-idealities arise from the limited precision and variations in the non-volatile memory (NVM) devices in the crossbars, leading to inaccurate dot-products and reduced inference accuracy \cite{sun2019impact, bhattacharjee2022examining}. Variation-aware Training (VAT) is widely employed to  improve robustness of neural networks against hardware non-idealities \cite{liu2015vortex, chakraborty2018technology}. Thus, online adaptation of pre-trained models to the specific hardware conditions \& non-idealities is imperative for resource-constrained edge platforms like IMC crossbars (see Fig. \ref{fig1}).

\begin{wrapfigure}{l}{0.55\columnwidth}
 \centering
 
\includegraphics[width=0.55\columnwidth]{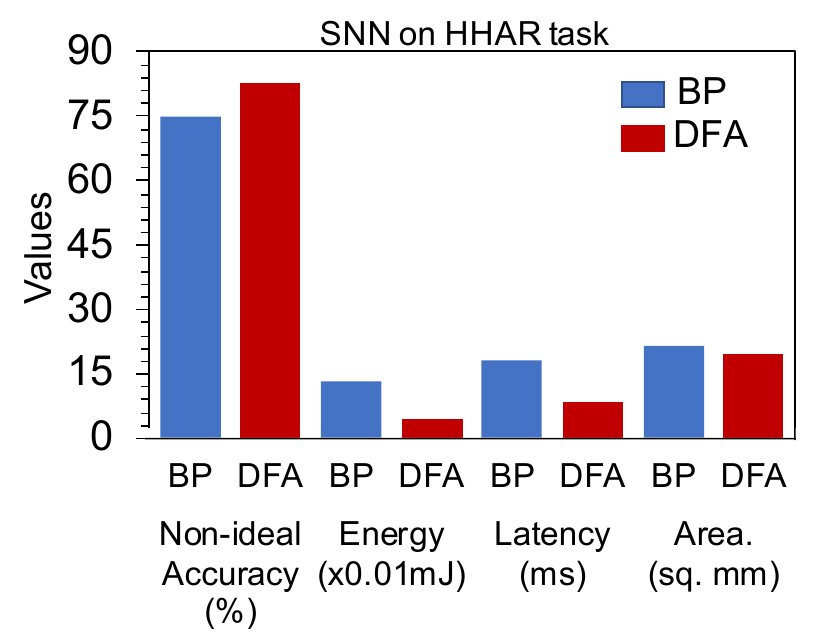}

\caption{Plot showing the advantages of DFA-based online SNN adaptation on an IMC platform over traditional BP.}

\label{fig:intro_res}
\end{wrapfigure} 

%The process of adapting SNNs on edge devices (performing VAT) in real-time should be facilitated with minimal energy, latency and area overheads, while maintaining algorithmic performance. However, traditional backpropagation (BP) based online adaptation on an analog IMC platform prior to inference entails the following pitfalls: (a) The layer-sequential backpropagation of gradients over multiple timesteps and epochs is latency-intensive. (b) BP on an IMC platform requires that the underlying crossbars be transposable \cite{peng2020dnn+, jin2021rehy} (details in Section \ref{sec:dfasim}), thus exacerbating the energy and area costs of the peripheral circuits that interface the crossbars. 

Adapting SNNs on edge devices with minimal energy, latency, and area overhead, while maintaining algorithmic performance, is paramount. However, traditional backpropagation (BP) for online adaptation on IMC platforms faces several challenges: (a) layer-by-layer gradient propagation across multiple timesteps is latency-intensive, and (b) BP requires transposable crossbars, thereby increasing the energy and area costs of peripheral circuits \cite{peng2020dnn+, jin2021rehy}. To address these limitations, we propose a Direct Feedback Alignment (DFA) strategy for online adaptation of pre-trained SNNs on RRAM crossbars, enabling robust inference. Unlike BP, DFA uses localized gradient learning to simultaneously fine-tune all SNN layers on hardware \cite{nokland2016direct}. The key contributions of our work are as follows:

%Addressing the above bottlenecks, we propose Direct Feedback Alignment (DFA) based online adaptation strategy for pre-trained SNNs deployed on RRAM crossbars for robust inference. Unlike BP, DFA entails a layerwise local gradient-based learning rule to simultaneously fine-tune the pre-trained weights of all the SNN layers on hardware \cite{nokland2016direct}. The key highlights of our work are as follows:

\begin{enumerate}

    %\item We implement our in-house evaluation engine called $DFA\_Sim$ to perform a holistic hardware cost analysis of our DFA-based online adaptation method against BP on RRAM crossbars. 

    \item Development of $DFA\_Sim$, our in-house evaluation engine for analyzing the hardware costs of DFA-based online adaptation compared to BP on RRAM crossbars.
    
    \item For hardware-realistic SNN accuracy evaluations using $DFA\_Sim$, we propose an accurate noise prediction model for RRAM devices in the crossbars using Gaussian Process Regression \cite{hossen2022data} with experimental data from a real IMC chip called NeuRRAM \cite{wan2022compute}.

    %\item We find that DFA-based adaptation of pre-trained SNNs on a human activity recognition (HAR) \cite{stisen2015smart} task incurs $2.1\times$ lower latency at $64.1\%$ lower energy \& $10.1\%$ lower area on RRAM crossbars than traditional BP-based adaptation, while achieving $7.55\%$ higher non-ideal inference accuracy (see Fig. \ref{fig:intro_res}). 

    \item Our experiments show that DFA-based SNN adaptation on a HAR task \cite{stisen2015smart} incurs $2.1\times$ lower latency, $64.1\%$ lower energy, $10.1\%$ lower area, and $7.55\%$ higher inference accuracy on RRAM crossbars compared to BP (see Fig. \ref{fig:intro_res}).

\end{enumerate}

\vspace{-5mm}

\section{Background}
\label{sec:background}

\begin{figure}[t]
    %\centering
    
    \includegraphics[width=\linewidth]{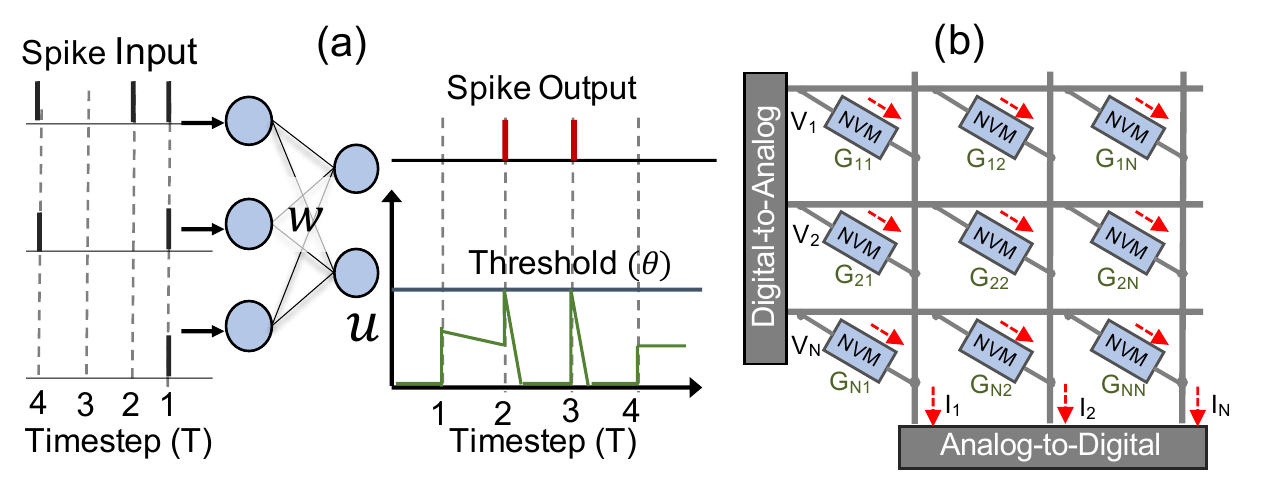}%

    %\captionsetup{justification=centering}
    \caption{(a) Pictorial representation of an SNN. (b) Pictorial representation of an N$\times$N crossbar.}
    \label{fig3}
    \vspace{-5mm}
\end{figure}

\textbf{Spiking Neural Networks:} As shown in Fig. \ref{fig3}(a), a key characteristic of SNNs is their use of a distinct neuronal activation function, typically the Leaky-Integrate-and-Fire (LIF) model, for temporal signal processing, in contrast to the ReLU activation commonly employed in Artificial Neural Networks (ANNs). The LIF neuron $i$, with its associated membrane potential $u_{i}^{t}$, integrates a series of spike inputs as follows:
\vspace{-2mm}
\begin{equation}
    u_i^t = \lambda  u_i^{t-1} + \sum_j w_{ij}o^t_j.
    \label{eq:LIF}
\vspace{-3mm}
\end{equation}

Here, $t$ stands for the timestep, $w_{ij}$ for weight connections between neuron $i$ and neuron $j$ and $\lambda$ denotes the leak factor. The  LIF neuron $i$ generates an output spike $o_i^{t}$ at the end of timestep $t$ if the membrane potential exceeds a threshold $\theta$:
\vspace{-2mm}
\begin{equation}
    o^{t}_i =
\begin{cases}
 1,          & \text{if $u_i^{t} >\theta$},  \\
    0
    & \text{otherwise.} 
\end{cases}
\label{eq:firing}
\end{equation}
\vspace{-4mm}
% \vspace{-1mm}

When the neuron fires, its membrane potential resets to zero. The integrate-and-fire mechanism of an LIF neuron leads to a non-differentiable function, which complicates the use of  backpropagation for training SNNs. To overcome this challenge, techniques such as Surrogate Gradient Learning or Backpropagation Through Time (BPTT) approximate the backward gradient function \cite{wu2018spatio}, enabling direct learning from spikes with fewer timesteps. Additionally, BPTT-based training of SNNs can be implemented using widely-used machine learning frameworks like PyTorch \cite{paszke2017automatic}.

Moreover, in line with prior work \cite{kim2022rate}, we use direct encoding to convert the input tensor into spike trains across a total of $T$ timesteps. For the final prediction, we run the inference over $T$ timesteps ($t=1,2,...,T$) and compute the average of the outputs from the SNN classifier.

\begin{figure}[b]
    \centering
    
    \includegraphics[width=0.8\linewidth]{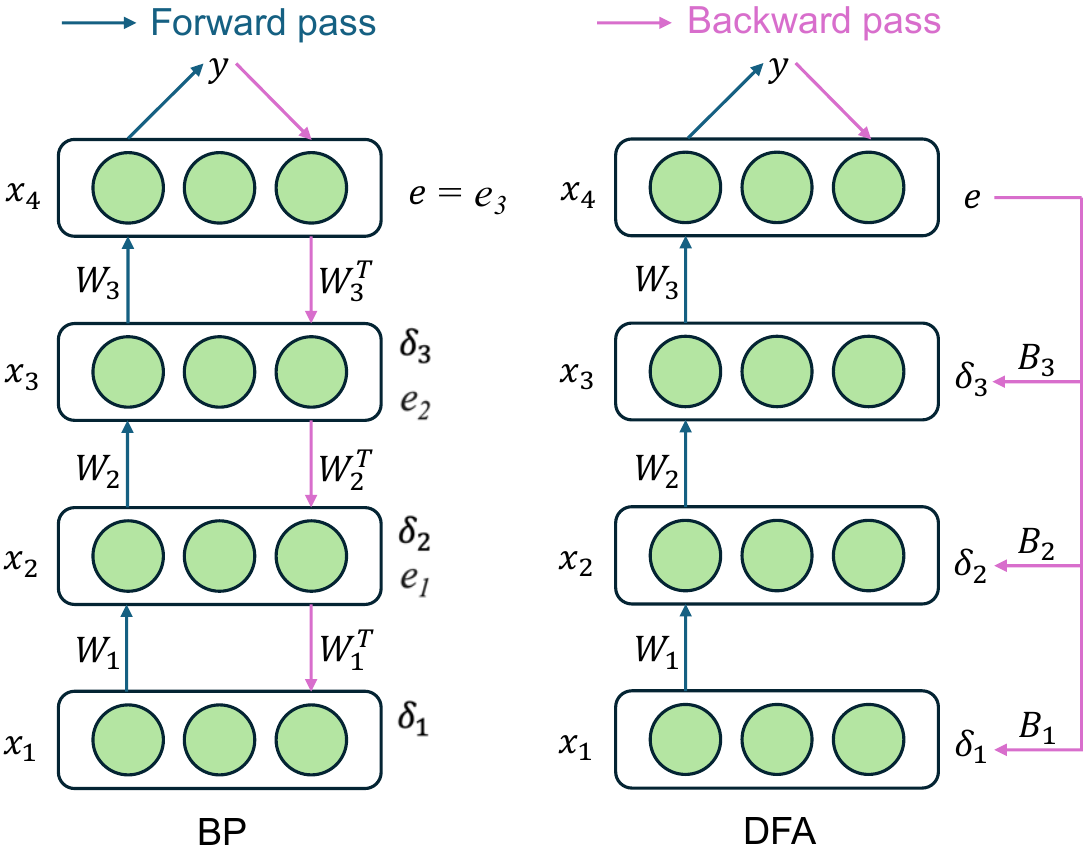}%

    %\captionsetup{justification=centering}
    \caption{Pictorial representations of BP (left) and DFA (right) learning.}
    \label{bpvsdfa}
\end{figure}

\begin{figure*}[t]
    \centering
    
    \includegraphics[width=0.85\linewidth]{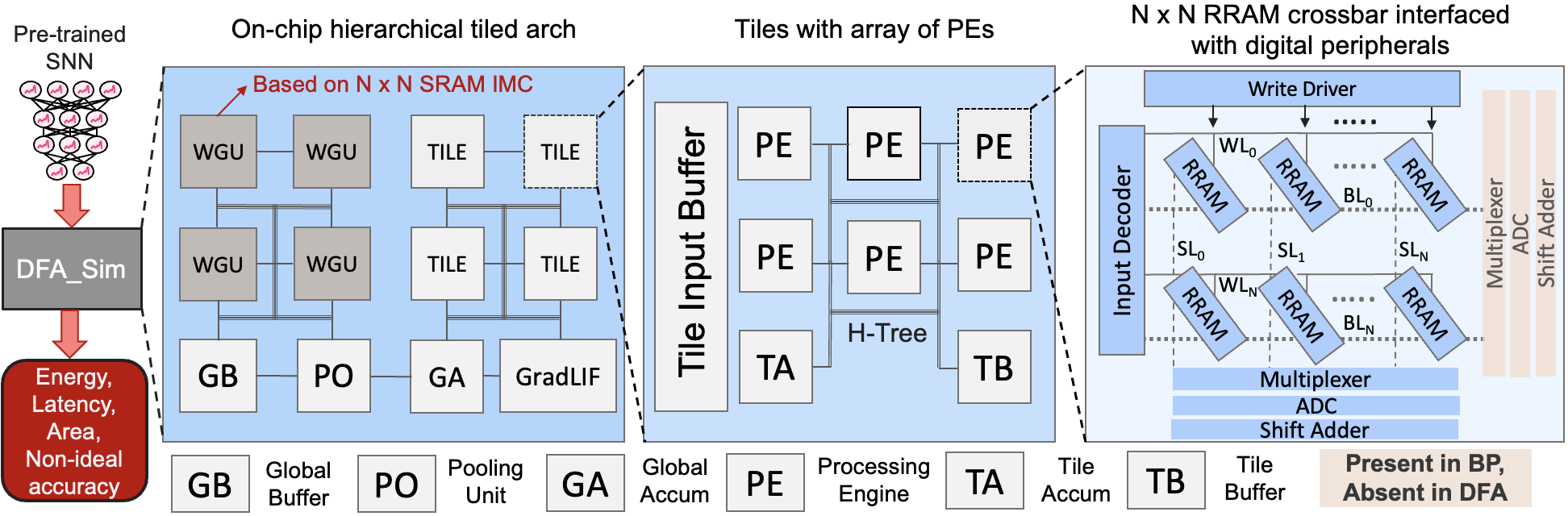}%

    %\captionsetup{justification=centering}
    \caption{Our proposed $DFA\_Sim$ engine. The hierarchical architecture consists of Tiles, Processing Engines (PEs) and RRAM crossbars. For DFA-based adaptation, the crossbars inside the PEs need not to be transposable. For BP-based adaptation the crossbars are transposable, thereby requiring extra periperals. This figure is for representation purpose only; actual number of tiles, PEs and crossbars may differ.}
    \label{dfa_sim}
    \vspace{-6mm}
\end{figure*}

\textbf{Analog IMC Crossbars:} Analog crossbars comprise of a 2D array of NVM devices,  interfaced with Digital-to-Analog Converters (DACs), Analog-to-Digital Converters (ADCs), and write circuits dedicated towards programming the NVM devices \cite{shanbhag2022comprehending, verma2019memory}. The SNN's spike inputs are encoded as analog voltages $V_i$ to each row of the crossbar by the DACs, while weights are programmed as NVM device conductances ($G_{ij}$) at the cross-points (we use RRAM as NVM devices in this work), as shown in Fig. \ref{fig3}(b). 

To emulate dot-product operations in an ideal N$\times$N crossbar during inference, input voltages interact with device conductances, generating currents according to Ohm's Law. Based on Kirchhoff's current law, the total output current sensed at each column $j$ by the ADCs is the sum of currents flowing through all devices, expressed as $I_{j(ideal)} = \Sigma_{i=1}^{N}{G_{ij} * V_i}$. However, in practical applications, the analog nature of computation introduces non-idealities, such as variations in non-volatile memory (NVM) devices \cite{sun2019impact, bhattacharjee2021neat}. Consequently, the net output current at each column $j$ deviates from the ideal value $I_{j(ideal)}$, leading to significant accuracy degradation in SNNs implemented on crossbars \cite{bhattacharjee2022examining, moitra2024when}.

\textbf{Direct Feedback Alignment (DFA):} Direct Feedback Alignment (DFA) is a recent learning approach designed to address some of the key bottlenecks of traditional BP \cite{nokland2016direct}. During the backward pass in DFA-based learning, as shown in Fig. \ref{bpvsdfa}, the feedback signals are aligned directly with the output errors $e$ by fixed, randomized feedback matrix connections ($B$). This method decouples the layerwise sequential process in BP to compute gradients for each layer $l$ ($\delta_l=e.B_l$) simultaneously, contrary to the gradient computation $\delta_l=e_{l}.W_l^T$ in BP. DFA, thus, enables parallel weight update of all layers by locally calculating gradients. We will see in Section \ref{sec:result} that the compatibility of DFA with analog IMC platforms makes it a promising solution to perform online noise adaptation of deployed SNNs in real-time.

\vspace{-3mm}

\section{$DFA\_Sim$: Hardware Evaluation Engine}

\label{sec:dfasim}

\textbf{Architectural Details:} $DFA\_Sim$ is a Python-based hardware evaluation engine to benchmark energy, latency \& area costs of DFA-based online adaptation of SNNs on a monolithic IMC chip built upon analog RRAM crossbars. As shown in Fig. \ref{dfa_sim}, it deploys SNN models on a hierarchical, weight-stationary tiled architecture, similar to SpikeSim \cite{moitra2023spikesim}. $DFA\_Sim$ features an array of interconnected tiles with global buffers, accumulators, LIF activation units, and pooling units implemented digitally. The global LIF activation unit (GradLIF), based on \cite{yin2022sata}, supports LIF operations in the forward pass and gradient calculations in the backward pass. Each tile contains 4 Processing Engines (PEs), input/output buffers, and accumulation modules for partial sums. Each PE includes 4 analog 256$\times$256 IMC crossbars using 4-bit RRAM devices \cite{wan2022compute} that perform dot-product operations, along with peripheral circuits such as input decoders, ADCs, write drivers, shift adders, etc. The PEs compute dot-products $x_l.W_l$ in the forward pass and $\delta_l = e.B_l$ during the backward pass (see Fig. \ref{bpvsdfa}) for a given SNN layer $l$. Weight gradients are calculated using dot-products $\delta_l.x_l$ in specialized digital SRAM-based Weight Gradient Units (WGUs) \cite{peng2020dnn+}. Global H-trees connect tiles and the global buffer, while local H-trees manage communication within each tile and PE.

%\textbf{Differences between Hardware for DFA and BP:} As shown in Fig. \ref{bpvsdfa}, BP computes the gradients $\delta_l=e_{l}.W_l^T$ in a layer-sequential manner. Thus, from a hardware perspective, BP requires transposable crossbars in the PEs to facilitate $x_l.W_l$ in the forward pass and $e_{l}.W_l^T$ during the backward pass. Transposable crossbars are  implemented by duplicating peripheral circuits (such as, ADCs, shift adders, etc.) on row-side of the arrays \cite{peng2020dnn+, wan2022compute}. DFA eliminates the requirement for transposable crossbars, thereby reducing the redundant area overhead for peripheral circuits in the PEs. However, DFA requires extra RRAM crossbars to perform feedback error signal calculation by programming them with random $B$ matrices, incurring additional area costs. Our analyses in Section \ref{sec:result} reveal that this additional area cost is compensated by the greater area savings at a PE-level by eliminating transposable crossbars and associated peripherals. Another interesting facet of using RRAM crossbars is that the intrinsic stochasticity of RRAM devices can be exploited to generate and store the random $B$ matrices in crossbars \cite{lin2019bayesian, malhotra2020exploiting}.

\textbf{Differences between Hardware for DFA and BP:} BP computes gradients $\delta_l=e_{l}.W_l^T$ layer-by-layer (see Fig. \ref{bpvsdfa}), requiring transposable crossbars in PEs to handle both $x_l.W_l$ in the forward pass and $e_{l}.W_l^T$ in the backward pass. This is achieved by duplicating peripheral circuits on the row-side of the crossbars \cite{peng2020dnn+, wan2022compute}. In contrast, DFA removes the need for transposable crossbars, reducing the significant area overhead of peripheral circuits. However, DFA requires additional RRAM crossbars for feedback error signal computation using random $B$ matrices. Section \ref{sec:result} shows that the area savings from eliminating transposable crossbars outweigh the extra cost of the feedback crossbars. Another interesting facet of using RRAM crossbars is that the intrinsic stochasticity of RRAM devices can be exploited to generate and store the random $B$ matrices \cite{lin2019bayesian, malhotra2020exploiting}.

\begin{figure*}[t]
    \centering
    
    \includegraphics[width=0.85\linewidth]{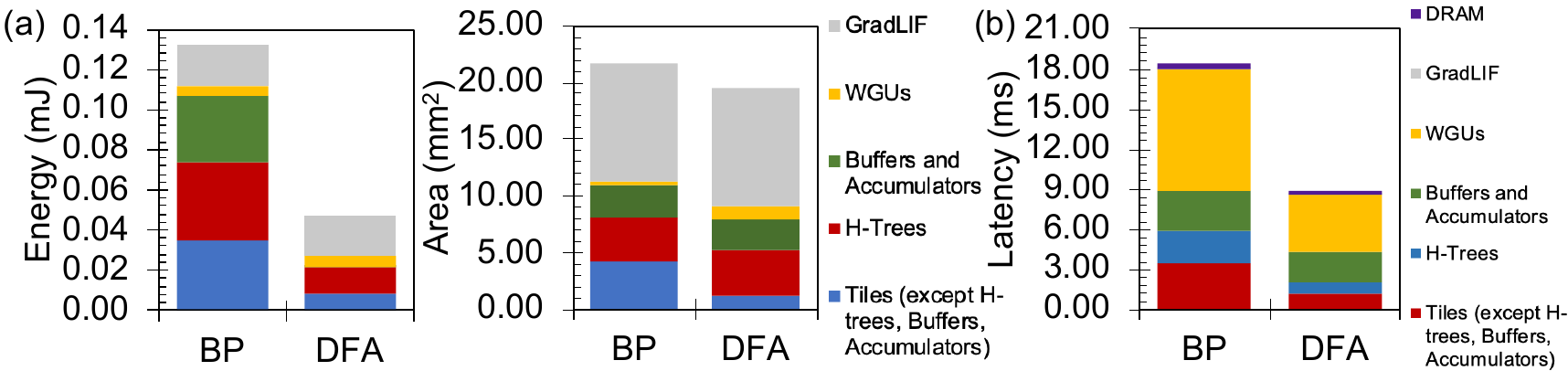}%

    %\captionsetup{justification=centering}
    \caption{Plots showing component-wise breakdown of--- (a) energy per epoch (left) and area (right) for online adaptation of SNN using HHAR dataset. (b) latency per epoch (left) for online adaptation of SNN using HHAR dataset.}
    \label{bar-charts}
    \vspace{-2mm}
\end{figure*}

\begin{table*}[t]
\Huge
\centering
\caption{Table comparing the performance of online adaptation of SNNs using DFA and BP methods using the $DFA\_Sim$ engine in terms of post-adaptation non-ideal test accuracy as well as energy, latency and area for online adaptation. The best results are \textbf{highlighted}.}
\label{tab:results}
\resizebox{.9\linewidth}{!}{
\begin{tabular}{@{}ccccccccccccc@{}}
\toprule
 &
   &
   &
   &
   &
  \multicolumn{2}{c}{\textbf{\begin{tabular}[c]{@{}c@{}}Non-ideal \\ HW Accuracy (\%)\end{tabular}}} &
  \multicolumn{2}{c}{\textbf{\begin{tabular}[c]{@{}c@{}}Energy \\ (mJ)\end{tabular}}} &
  \multicolumn{2}{c}{\textbf{\begin{tabular}[c]{@{}c@{}}Latency \\ (ms)\end{tabular}}} &
  \multicolumn{2}{c}{\textbf{\begin{tabular}[c]{@{}c@{}}Area \\ (mm$^2$)\end{tabular}}} \\ \midrule
\textbf{\begin{tabular}[c]{@{}c@{}}Timesteps \\ (T)\end{tabular}} &
  \textbf{\begin{tabular}[c]{@{}c@{}}SNN MLP \\ Architecture\end{tabular}} &
  \textbf{\begin{tabular}[c]{@{}c@{}}Task \\ (Dataset)\end{tabular}} &
  \textbf{\begin{tabular}[c]{@{}c@{}}Pre-trained \\  FP32 Accuracy (\%)\end{tabular}} &
  \textbf{\begin{tabular}[c]{@{}c@{}}Pre-adaptation \\ Non-ideal \\ HW Accuracy (\%)\end{tabular}} &
  \textbf{BP} &
  \textbf{DFA} &
  \textbf{BP} &
  \textbf{DFA} &
  \textbf{BP} &
  \textbf{DFA} &
  \textbf{BP} &
  \textbf{DFA} \\ \midrule
5   & 784-512-256-128-64-10 & Fashion MNIST & 88.55 & 61.92 & 83.55 & \textbf{85.12} & 0.0181 & \textbf{0.0072} & 1.9 & \textbf{0.947} & 23.1 & \textbf{20.1} \\
128 & 9-128-64-32-6         & UCI-HAR       & 89.68 & 57.22 & 84.93 & \textbf{85.61} & 0.0979 & \textbf{0.0316} & 15.7 & \textbf{8.24} & 21.7 & \textbf{19.5} \\
100 & 6-256-128-64-6        & HHAR          & 90.77 & 39.6  & 75.0  & \textbf{82.55} & 0.132 & \textbf{0.0474} & 18.1 & \textbf{8.67} & 21.7 & \textbf{19.5} \\ \bottomrule
\end{tabular}}
\vspace{-4mm}
\end{table*}

\begin{table}[t]
\Huge
    \caption{$DFA\_Sim$ Hardware Parameters\label{tab:dfa_sim}}
    \centering
    \resizebox{0.65\linewidth}{!}{
    \begin{tabular}{cc}
    \toprule
    \textbf{Parameters} & \textbf{Data} \\ 
    \midrule
     Technology & 32 nm CMOS \\
     Operating Frequency & 1 GHz\\
     Global/Tile/PE buffer & 16 KB/4 KB/1 KB \\
     PEs per Tile, Crossbars per PE & 4 , 4 \\
     Crossbar size & 256$\times$256 \\
     RRAM Device Precision & 4 bits \\
     $[G_{min}, G_{max}]$ & [1$\mu S$, 10$\mu S$]\\
     Spike-decoder Precision & 1 bit \\
     ADC Precision & 4 bits \\
    \bottomrule
     \end{tabular}}
     \vspace{-5mm}
\end{table}

\textbf{Mapping SNNs:} For SNN weight mapping, we adopt the standard approach proposed in SpikeSim \cite{moitra2023spikesim}, assuming that no two layers of an SNN can be mapped onto the PEs within a single tile. To enable DFA, we allocate an additional tile to store random $B$ matrices in the RRAM crossbars. Given that this work focuses on tasks like Human Activity Recognition (HAR) that are executed on highly resource-constrained hardware, simpler SNN architectures such as multi-layer perceptrons (MLPs) are preferred (see Table \ref{tab:results}) \cite{li2023efficient}. Due to the smaller scale of the SNN architectures and the smaller size of the $B$ matrices in DFA compared to $W^T$ in BP, all $B$ matrices can easily fit within a single tile (comprising 16 RRAM crossbars of size 256$\times$256) and we consider duplication of $B$ matrices in the crossbars to accelerate batch-training.

\begin{wrapfigure}{l}{0.56\columnwidth}
 \centering
 
\includegraphics[width=0.56\columnwidth]{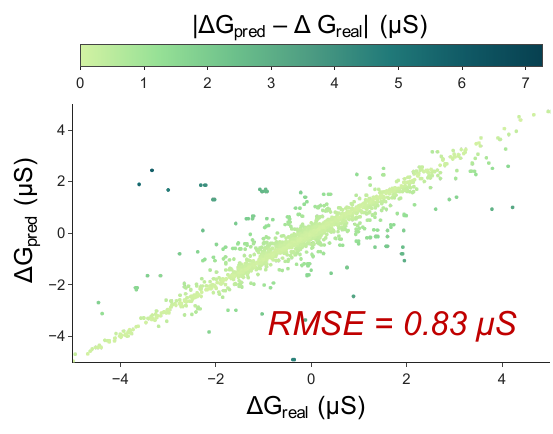}
\caption{Plot between $\Delta G_{real}$ (extracted from NeuRRAM data \cite{wan2022compute}) and $\Delta G_{pred}$.}

\label{fig:gpr_result}
\vspace{-3mm}
\end{wrapfigure}    

\textbf{Integrating Realistic RRAM Noise Model:} $DFA\_Sim$ is equipped with an accurate noise model for RRAM devices that predicts non-ideal conductance $G_{non-ideal}$ from the ideal conductance $G_{ideal}$. SNN weights are first mapped to RRAM conductances in the range $G_{ideal}\epsilon[G_{min}, G_{max}]$ with 4-bit precision. However, when dot-product operations are carried out in the crossbars, the conductances suffer from non-idealities stemming from the RRAM device variations. We use Gaussian Process Regression (GPR) to train our RRAM noise prediction model with experimental data acquired from a real IMC chip called NeuRRAM \cite{wan2022compute} and use it to predict the noise $\Delta G=G_{non-ideal}-G_{ideal}$ injected into the programmed RRAM conductances. Before noise-modelling with GPR using the GPyTorch package, data acquired from the NeuRRAM chip was cleaned to select conductances in the range of $G_{min}=1\mu S$ to $G_{max}=10\mu S$. Thereafter, the raw data from the chip was randomly divided into training (80\%, 6920 samples) and testing (20\%, 1730 samples) datasets to train the noise prediction model for 100 epochs. Fig.\ref{fig:gpr_result} shows the fitting results with the testing dataset. We find good agreement between the ground-truth $\Delta G_{real}$ and the predicted $\Delta G_{pred}$ with an $RMSE=0.83\mu S$.

\vspace{-3mm}

\section{Results and Discussion}
\label{sec:result}

%\textbf{Experimental Setup:} For all our experiments, we use BPTT-trained SNN MLPs on two HAR tasks (UCI-HAR \cite{anguita2013public} \& HHAR \cite{stisen2015smart}) and one image classification task (Fashion MNIST \cite{xiao2017fashion}) as shown in Table \ref{tab:results}. UCI-HAR contains instances collected from 30 subjects involving 6 different activities--- walking, walking upstairs, walking downstairs, sitting, standing, and lying. The sensors are accelerometer and gyroscope (sampling rate of 50 Hz) from Samsung Galaxy SII. HHAR contains instances collected from 9 subjects involving 6 daily activities--- biking, sitting, standing, walking, stair up, and stair down. The sensors are accelerometers from 8 smartphones and 4 smart watches (sampling rate from 50 to 200 Hz). Fashion MNIST consists of grayscale images of size 28$\times$28 from 10 classes. The pre-trained SNNs are subjected to online adaptation on non-ideal RRAM crossbars for 25 epochs. We use the $DFA\_Sim$ engine to estimate the hardware-realistic energy, latency, area and inference accuracy metrics on performing DFA-based online adaptation and compare against BP-based online adaptation. Here, all the energy \& latency values are calculated for one epoch of online adaptation. The training batch size is fixed at 50. The details of hardware setup for $DFA\_Sim$ are shown in Table \ref{tab:dfa_sim}. 

\textbf{Experimental Setup:} We conducted experiments using BPTT-trained SNN MLPs on two human activity recognition (HAR) tasks (UCI-HAR \cite{anguita2013public} and HHAR \cite{stisen2015smart}), and one image classification task (Fashion MNIST \cite{xiao2017fashion}), as summarized in Table \ref{tab:results}. UCI-HAR includes data from 30 subjects performing six activities (walking, walking upstairs, walking downstairs, sitting, standing, and lying) using accelerometer and gyroscope sensors from a Samsung Galaxy SII (50 Hz sampling rate). HHAR involves data from 9 subjects performing six daily activities (biking, sitting, standing, walking, stair up, and stair down) using accelerometers from 8 smartphones and 4 smartwatches (sampling rates between 50 and 200 Hz). Fashion MNIST consists of 28$\times$28 grayscale images from 10 classes. The pre-trained SNNs were adapted online on non-ideal RRAM crossbars for 25 epochs. We used the $DFA\_Sim$ engine to estimate energy, latency, area and inference accuracy for DFA-based online adaptation and compared it with BP-based adaptation. Energy and latency were calculated per epoch, with a fixed training batch size of 50. Hardware details for $DFA\_Sim$ are listed in Table \ref{tab:dfa_sim}.

%From Fig. \ref{bar-charts}(a), we find that DFA-based online adaptation on the HHAR task leads to $64.1\%$ reduction in training energy, mainly attributed to the reduced tile-level computation \& H-tree data communication costs. Please note that both BP and DFA incur a constant DRAM access energy expenditure (17.7 $mJ$ for the HHAR task) that is not included in Fig. \ref{bar-charts}(a) and Table \ref{tab:results}. We also find that DFA leads to an overall $10.1\%$ decrease in area compared to BP. This is mainly attributed to the decrease in tile area due to the elimination of transposable crossbars and associated peripherals. To facilitate simultaneous gradient computations for all layers, DFA requires an additional tile to process $\delta_l=e.B_l$ operations and a larger WGU area than BP to compute all weight updates together. However, the tile area saved ($13.4\%$) by eliminating transposable crossbars compensates for the marginal increase ($3.29\%$) in WGU area. Fig. \ref{bar-charts}(b) shows $2.1\times$ overall speedup achieved by DFA by simultaneously processing all SNN layers, mainly by reducing latency at tile, WGU and data communication-levels. 

From Fig. \ref{bar-charts}(a), we observe that DFA-based online adaptation for the HHAR task results in a $64.1\%$ reduction in training energy, primarily due to lower tile-level computation and H-tree data communication costs. Note that both BP and DFA incur a constant DRAM access energy cost (17.7 $mJ$ for the HHAR task), which is not shown in Fig. \ref{bar-charts}(a) or Table \ref{tab:results}. Additionally, DFA reduces the overall area by $10.1\%$ compared to BP, largely because of the elimination of transposable crossbars and their associated peripherals. While DFA requires an extra tile for the $\delta_l = e.B_l$ operations and a larger WGU area to handle simultaneous weight updates, the $13.4\%$ area saved by removing transposable crossbars offsets the $3.29\%$ increase in WGU area. Fig. \ref{bar-charts}(b) also demonstrates that DFA achieves a $2.1\times$ overall speedup by processing all SNN layers concurrently, primarily by reducing latency at the tile, WGU, and data communication levels.

Table \ref{tab:results} presents overall results to underscore the efficacy of DFA-based online adaptation of SNNs in real-time. While naively deploying SNNs on the IMC platform significantly reduces their performance ($\sim27-51\%$ loss in accuracy), online adaptation with DFA can restore their performance by reducing the accuracy losses to $\sim3-8\%$, compared to the FP32 software baseline. In fact, DFA leads to better performance ($\sim1-8\%$ higher non-ideal accuracy) than traditional BP-based online adaptation. This is because the layer-sequential gradient propagation in BP results in error accumulation due to non-idealities affecting the SNN weights. However, as DFA decouples gradient computations at a given SNN layer from its predecessors, error accumulation is eliminated. Furthermore, DFA achieves $\sim60-68\%$ lower energy at $\sim10-13\%$ lower area and $\sim2\times$ lower latency than BP.

\vspace{-3mm}

\section{Conclusion}
\label{sec:conclusion}

To the best of our knowledge, this work for the first time proposes DFA as a low-cost and efficient method for online adaptation of pre-trained SNNs on resource-constrained and non-ideal edge devices. Our in-house $DFA\_Sim$ engine highlights the significant energy, area, and latency benefits of DFA over traditional BP for real-time learning on an RRAM-based IMC platform. Furthermore, with a realistic RRAM noise prediction model integrated with $DFA\_Sim$, we show SNNs adapted using DFA to achieve better non-ideal accuracy compared to BP. These findings underscore the potential of DFA-based online adaptation for advancing low power, spike-based analytics in wearable and edge computing applications.

% \scriptsize
% \section*{Acknowledgement}This work was supported in part by CoCoSys, a JUMP2.0 center sponsored by DARPA and SRC, the National Science Foundation (CAREER Award, Grant \#2312366, Grant \#2318152), and the DoE MMICC center SEA-CROGS (Award \#DE-SC0023198).
% \normalsize

\bibliographystyle{IEEEtran}

\bibliography{reference}

\end{document}